\documentclass[floatfix,showpacs,amsmath,amssymb,letterpaper,groupaddresses,superscriptaddress]{article}
\usepackage{graphicx}
\usepackage{latexsym}
\usepackage{amsmath}
\usepackage{amssymb}
\usepackage{epsfig}
\usepackage{graphicx}
\usepackage{graphicx}
\usepackage{amssymb,amsmath,times}
\usepackage{hyperref}
\usepackage{color}

\def\be{\begin{equation}}
\def\ee{\end{equation}}
\def\ba{\begin{eqnarray}}
\def\ea{\end{eqnarray}}
\def\la{\langle}
\def\ra{\rangle}

\def\h{\hskip 1cm}

\usepackage{epsfig}
\usepackage{graphicx}
\begin{document}
\begin{center}
{\Large \bf Systematic study of the completeness of two-dimensional classical $\phi^4$ theory\\
}
\vspace{1cm}
Mohammad Hossein Zarei\footnote{email:mzarei92@shirazu.ac.ir}\h
Yahya Khalili\footnote{email:yahyakhalili@yahoo.com}\\
\vspace{5mm}
\vspace{1cm} Physics Department, College of Sciences, Shiraz University, Shiraz 71454, Iran\\
\end{center}
\vskip 3cm
\begin{abstract}
The completeness of some classical statistical mechanical (SM)
models is a recent result that has been developed by quantum
formalism for the partition functions. In this paper, we consider a 2D classical $\phi^4$ filed theory whose completeness
has been proved in [V. Karimipour and et al, Phys. Rev. A 85, 032316]. We give a new and general systematic proof for the completeness of such a model where, by a few simple steps, we show how the partition function of an arbitrary classical field theory can be derived from a 2D classical $\phi^4$ model. To this end, we start from various classical field theories containing models on
arbitrary lattices and also $U(1)$ lattice gauge theories. Then we
convert them to a new classical field model on a non-planar
bipartite graph with imaginary kinetic terms. After that, we show
that any polynomial function of the field in the corresponding
Hamiltonian can approximately be converted to a $\phi^4$ term by adding enough
numbers of vertices to the bipartite graph. In the next step, we
give a few graphical transformations to convert the final
non-planar graph to a 2D rectangular lattice. We also show that
the number of vertices which should be added grows polynomially
with the number of vertices in the original model.
\end{abstract}
\section{Introduction}
Meeting between quantum information theory and classical
statistical mechanics is a recent field of research that has
received a great deal of interest during the past decade \cite{13,
14, 15, 16, 17, 18, 19}. Specially, one of the interesting results
of these connections has led to a kind of the unification in
statistical mechanics which has been called completeness \cite{4}. Since
all thermodynamics properties of a model can be derived from the
partition function \cite{kar}, the main idea is related to unify the
partition functions of various SM models in the partition function
of a complete model. In this way, there is a specific model
Hamiltonian where the partition function of such a
model, in different regimes of coupling constants, can generate the
partition functions of other SM models. Furthermore, in a recent paper \cite{21}, a new concept of the universal models has also been introduced where, by a bridging to computer science, it has been shown that 2D Ising model in presence of magnetic field with real coupling constants is a universal model. It means that the physics of all classical models with discrete or continuous degrees of freedom are reproduced in the low-energy sector of a 2D Ising model. By such a definition, a complete model is also a specific kind of a universal model.

The main key to find a complete model has emerged from a quantum formalism for the partition functions. In \cite{5}, authors showed that the partition function of a spin model can be written as the inner product of two quantum states, see also \cite{50}. By quantum formalism, the
first complete model was introduced in an interesting paper
\cite{4} where the authors showed that the partition function of the
2D Ising model with complex interactions and magnetic
field is equal to the partition function of any SM models. In fact, one can imagine
the phase diagram of a 2D Ising model with the magnetic field where it covers all universality classes.

In fact, the completeness of 2D Ising model is related to the universality of cluster states for measurement-based quantum computation \cite{6, 7, 8}. A similar approach also led to other complete models like 4D $Z_2$ lattice gauge theory \cite{9, 10}. An extension of above results to statistical models with continuous degrees of freedom has also led to completeness of 4D $U(1)$ lattice gauge theory \cite{11}
and 2D classical $\phi^4$ field theory \cite{12}. 

The above results of the completeness are actually some proofs of principle. In fact, it is proved that corresponding to each SM model, there is principally a suitable pattern of couplings for the complete model in the sense that the corresponding partition function is equal to the partition function of the original model. However, it is not a simple task to find a specific pattern of couplings related to each SM model. To this reason, re-considering certain complete models has attracted much attention in recent years. In particular, in \cite{20}, authors have given an algorithmic proof for the completeness of two-dimensional Ising model where one might be able to find the complete model corresponding to each SM model by following a few graphical transformations.

With a view to re-considering different complete models, here we study the 2D $\phi^4$ field theory. In \cite{12}, a proof of principle has been given for the completeness of such a model in SM models with continuous variables. The main goal of our paper is to give a systematic proof for the completeness of 2D $\phi^4$ theory. In fact we show how the partition function of any classical filed model can be derived from a 2D $\phi^4$ model by following a few simple steps. To this end, we start from a discrete version of a general classical field theory on an arbitrary lattice where the corresponding Hamiltonian contains arbitrary polynomial functions of fields. Then we convert such models to a new classical field theory with imaginary kinetic terms on a non-planar bipartite graph. We also show a similar result for $U(1)$ lattice gauge theory. After that, we convert all polynomial functions of the final model to a $\phi^4$ term by adding enough numbers of vertices with suitable coupling terms to the bipartite graph.

The above transformations on the original graph actually leads to a complex graph where degree of vertices can be generally very large. Therefore, we give a graphical transformation to reduce degrees of all vertices to four. The final graph is also a non-planar graph and we give another graphical rule to convert it to a planar graph. In this way, by combination of the above transformations, one will be able to convert the original model to a classical $\phi^4$ theory on a 2D rectangular lattice with suitable coupling terms on each vertex. 
In this way, by following a few graphical transformations, one will be able to find coupling terms of a 2D $\phi^4$ theory corresponding to any other classical field theory and also $U(1)$ lattice gauge theories. Furthermore, since we only use some mathematical lemmas to prove the above transformations, it will clarify the previous result that had been derived by quantum formalism for other communities of physics as well as quantum information theorists.

The structure of this paper is as follows: In section (\ref{sec1})
we briefly review the definitions of the general classical field
theories and also $U(1)$ lattice gauge theories. In section
(\ref{sec2}) we give a simple transformation on the partition
function of the above models to convert them to a unique form with
a simple kinetic term in the Hamiltonian. In section (\ref{sec3}),
we consider other terms of the Hamiltonian which can be arbitrary
polynomial functions and give a transformation to reduce each
polynomial function to a $\phi^4$ term. In section (\ref{sec4}),
we consider how to decrease dimensions of initial models to two
and then in section (\ref{sec5}), we give the last transformations
to match completely the model to a 2D square lattice. Finally, in
section (\ref{sec6}) we will find the complete 2D $\phi^4$ field
theory corresponding to the original model and discuss on the efficiency of our transformations.
\section{A brief review of classical field theories and $U(1)$ lattice gauge theories}\label{sec1}
In this section, we briefly review the definitions of two categories of field theories, classical field theories and $U(1)$ lattice gauge theories. These models belong to two important symmetrical categories containing global and local symmetries.

Classical field theories have important theoretical applications in different fields of physics. On the one hand, they are important models for studying different ideas in statistical mechanics \cite{f0, f1, f2, f3} and on the other hands they are useful for studying high-temperature behavior of quantum field theories \cite{q5, q6} and for studying non-perturbative dynamics of low-momentum models in non-equilibrium quantum field theory \cite{q1, q2, q3, q4}.

In this paper, we will deal with a general form of classical field theories. A classical field theory is defined by a Hamiltonian which can be a polynomial function of fields and derivations of fields in the following general form:
\begin{equation}
H=\int D\phi ~h(\partial \phi , \partial^2 \phi , ...(\partial \phi)^2, ..., \phi , \phi^2 ,...,\phi \partial \phi, ...).
\end{equation}
Here we consider only a specific form of such models which are more well-known in the following form:
\begin{equation}
H=-\int D\phi \{V(\partial \phi)+W(\phi)\},
\end{equation}
\begin{figure}[t]
\centering
\includegraphics[width=10cm,height=4.5cm,angle=0]{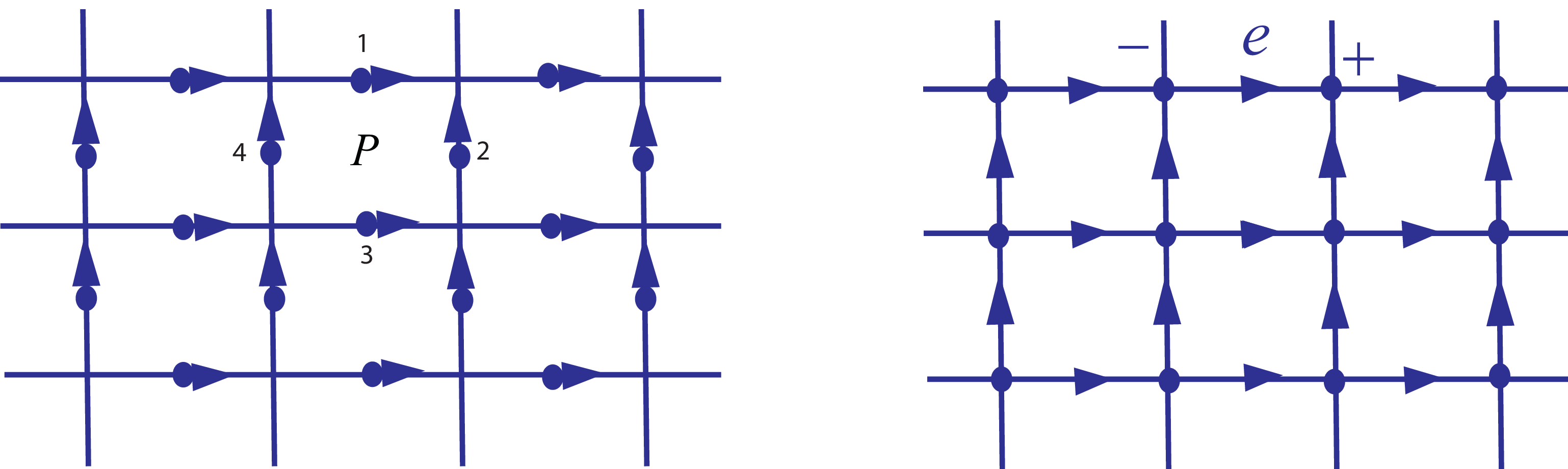}
\caption{(Color online) Left: An oriented lattice for $U(1)$ lattice gauge theory where field variables live in edges of the lattice. A plaquette of the lattice include four field variables which are denoted by numbers 1 to 4. right: An oriented lattice for classical field theory where field variables live in vertices of the lattice. Two variables in two endpoints of each edge are denoted by $+$ and $-$.} \label{fig0}
\end{figure}
where $V(\partial \phi)$ and $W(\phi)$ are some polynomial functions of $\partial \phi$ and $\phi$, respectively. Although such a model is defined on the continuous spacetime, there is also a discrete version of that on the discrete spacetime. To this end, the continuous spacetime is approximated by a lattice where continuous variables $\phi_i$ live in the vertices of the lattice, see Figure (\ref{fig0}, right) as a sample lattice. After discretization, the Hamiltonian is converted to a new one on an oriented lattice in the following form:
\begin{equation}\label{q0}
H=-\sum_{e}V_{e}(\phi_{+} - \phi_{-})-\sum_{i}W_{i}(\phi_i),
\end{equation}
where $e$ refers to the edges of the lattice and $\phi_{+}$ and $\phi_{-}$ refer to two variables which live in two endpoints of the edge $e$, see Figure(\ref{fig0}, right). $V_{e}$ and $W_i$ refer to an arbitrary polynomial function, and they can be in different forms for various edges and vertices of the lattice. We call $V_e$ and $W_i$ edge term and vertex term of the Hamiltonian, respectively.

Another kind of important field theories with local symmetries are $U(1)$ lattice gauge theories which are known as a discrete version of electrodynamics \cite{u0} and are also generalizations of Wegner's Ising gauge theories \cite{u1, u2, u3, u4}. While there are two different kinds of such models which are called compact and non-compact lattice gauge theories \cite{c1, c2, c3}, in this paper, we will deal with the compact one where the degrees of freedom are directly elements of $U(1)$ group.

A $U(1)$ lattice gauge theory is defined on a rectangular lattice which is considered as a discretized spacetime in an arbitrary dimension where continuous variables live in edges of the lattice, see Figure (\ref{fig0}, left) as a sample lattice. For arbitrary directions on all edges of the lattice like Figure(\ref{fig0}, left), the Hamiltonian of this model is defined in the following form:
\begin{equation}\label{q1}
H=-\sum_{p}J_p Cos(\phi_1 -\phi_2 -\phi_3 + \phi_4),
\end{equation}
where $J_p$ refers to coupling constants corresponding to each plaquette of the lattice which is denoted by $"p"$ and sign of variables $\phi_i$ is chosen according to directions of corresponding edges when we traverse a plaquette in the clockwise direction.

\section{Transformation to a simple kinetic term}\label{sec2}
In this section, we consider the partition function of a discrete version of the classical field theories and $U(1)$ lattice gauge theories. We show that a large set of field theories and also $U(1)$ lattice gauge theories can be unified in a new model with a simple kinetic term.

At the first, consider the discrete version of a classical field model on an arbitrary lattice with $N$ vertices where continuous variables $\phi_i$ live in the vertices of the lattice in the form of (\ref{q0}). The partition function of such a model is written in the following form:
\begin{equation}\label{a0}
\mathcal{Z}=\int D \phi~ e^{\sum_{e}V_{e}(\phi_{+} - \phi_{-})+\sum_{i}W_{i}(\phi_i)},
\end{equation}
\begin{figure}[t]
\centering
\includegraphics[width=10cm,height=3.5cm,angle=0]{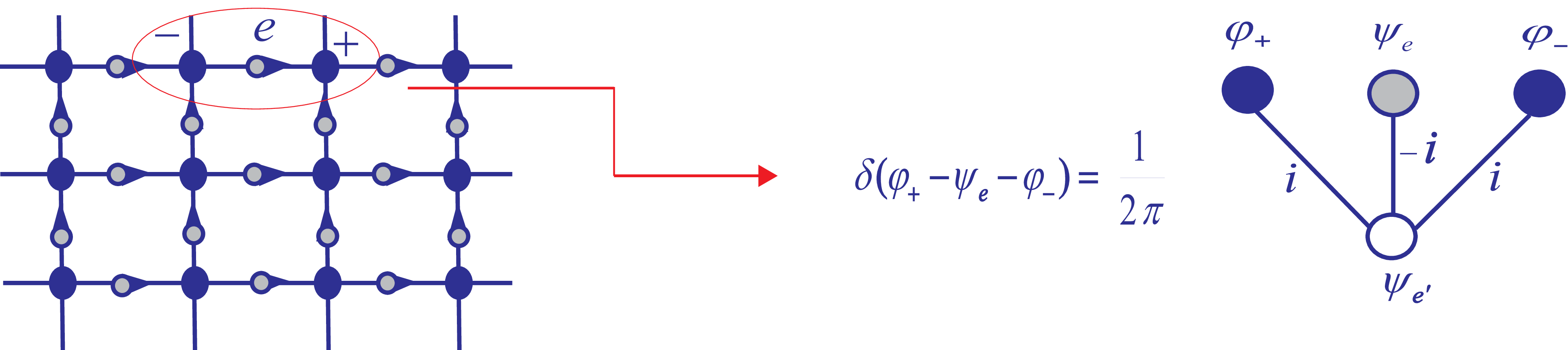}
\caption{(Color online) Left: New field variables are inserted in any edges of the lattice and are denoted by gray circles. Right: A constraint between the edge variable $\psi_e$ and vertex variables $\psi_+$ and $\psi_-$ is applied by a delta function which is equal to the partition function of a model corresponding to right-hand graph .} \label{fig1}
\end{figure}
where $D \phi$ refers to product of $d\phi_1 d\phi_2 ... d\phi_N$ and we absorb coefficient $\beta=\frac{1}{k_{B}T}$ in the Hamiltonian for all relations in this paper.

In the first step, consider an edge of the lattice which is denoted by $e$. We define new continuous variables on any edges of the lattice as $\psi_{e}= \phi_{+} -\phi_{-} $ and replace them in the partition function in relation (\ref{a0}). Due to this change of variable, it is necessary to add many delta functions in the partition function. In fact, new variables $\psi_{e}$ are not independent variables, and there are many constraints on them. Since $\psi_{e}=\phi_{+} -\phi_{-}$ we conclude that $\phi_{+} -\psi_{e} -\phi_{-}=0$. We can apply such a constraint in the relation of the partition function by an integral of a delta function in the form of $\int d\psi_{e}\delta(\phi_{+} -\psi_{e} -\phi_{-})$. It is clear that there are several constraints corresponding to all edges of the lattice where after the change of variables, several delta functions would add to satisfy the constraints. Therefore the partition function in relation (\ref{a0}) is converted to a new one in the following form:
\begin{equation}\label{a1}
\mathcal{Z}=\int D\psi D \phi ~e^{\sum_{e}V_{e}(\psi_{e})+\sum_{i}W_{i}(\phi_i)}\prod_{e}\delta(\phi_{+} -\psi_{e} -\phi_{-}),
\end{equation}
where $D\psi$ refers to product of $\prod_{e} d\psi_{e}$. 

In the next step, we use a simple relation for the definition of a delta function in the following form:
\begin{equation}
\delta(\phi_{+} -\psi_{e} -\phi_{-})=\frac{1}{2\pi}\int e^{i\psi_{e'}(\phi_{+} -\psi_{e} -\phi_{-})}d\psi_{e'},
\end{equation}
where we define new continuous variables $\psi_{e'}$ corresponding to all edges of the lattice. This relation is graphically shown in Figure (\ref{fig1}). If we use the above relation instead of delta functions in the partition function, we will have:
\begin{equation}\label{a1}
\mathcal{Z}=\int D\psi D \phi~ e^{\sum_{e}V_{e}(\psi_{e})+\sum_{i}W_{i}(\phi_i)}\prod_{e}\int d\psi_{e'}~ e^{i\psi_{e'} (\phi_+ -\psi_{e} -\phi_{-})}.
\end{equation}
And after re-writing the partition function, it will be in the following form:
\begin{equation}\label{a1}
\mathcal{Z}=\frac{1}{(2\pi)^E}\int \prod d\psi_{e} \prod d\phi_i \prod d\psi_{e'}~ e^{i\sum_{e}\psi_{e'} (\phi_+ -\psi_{e} -\phi_{-})+ \sum_{e}V_{e}(\psi_{e})+\sum_{i}W_{i}(\phi_i)},
\end{equation}
\begin{figure}[t]
\centering
\includegraphics[width=10cm,height=4.5cm,angle=0]{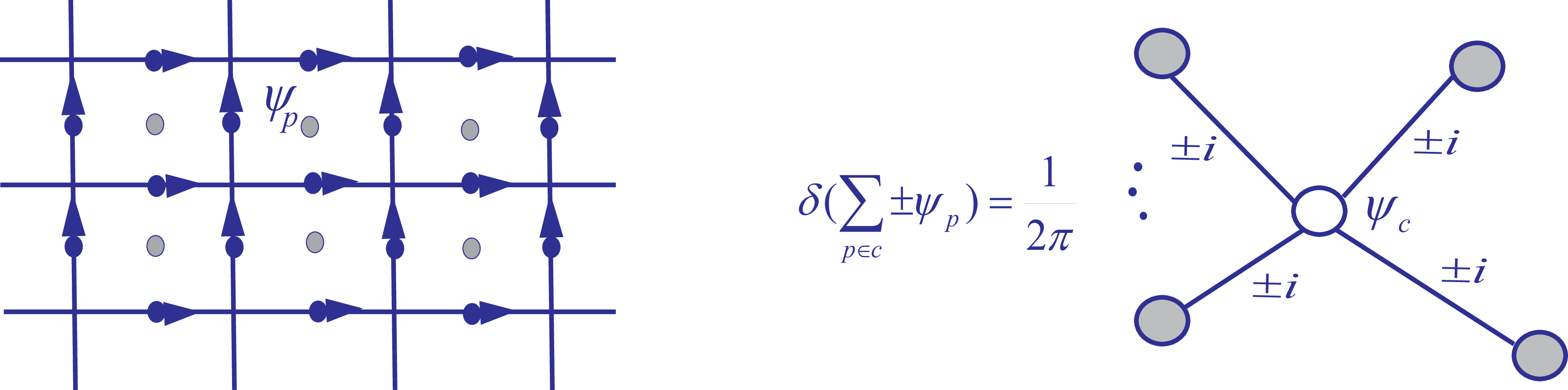}
\caption{(Color online) Left: New field variables are inserted in any plaquettes of the lattice and are denoted by gray circles. Right: Related to the structure of the lattice there are constraints between new plaquette variables $\psi_p$ which are applied by delta functions. Each delta function as $\delta(\sum \pm \psi_p)$ is equal to the partition function of a model corresponding to the right-hand graph. } \label{fig1-0}
\end{figure}
where $E$ is the number of edges of the lattice. The above
relation shows that the partition function of the initial model on
the original lattice is converted to the partition function of a new
model where we have added new variables $\psi_{e}$ and $\psi_{e'}$
corresponding to all edges of the lattice. In Figure (\ref{fig1})
we denote the vertex corresponding to $\psi_{e'}$ by light circles
and other vertices by blue circles where a white vertex is
connected to three blue vertices. In this way, if we apply such
transformations corresponding to all edges of the initial lattice,
we will have a bipartite graph where the partition function of new
model corresponds to a Hamiltonian in the following general form:
\begin{equation}
H=\sum_{e,i}\pm i\psi_{e'}\phi_i +i\psi_{e'} \psi_e -\sum_{i}W_{i}(\phi_i)-\sum_{e}V_{e}(\psi_e),
\end{equation}
where $+i$ and $-i$ correspond to vertex variables which had already been denoted by $\phi_{+}$ and $\phi_{-}$, respectively. If we use a simple identity that $ixy=-\frac{i}{2}[(x-y)^2 - x^2 -y^2]$, we can conclude that the new Hamiltonian is a new field theory which has a simple kinetic term as $\frac{i}{2}(\phi_i -\phi_j )^2$ instead of the edge terms $V_e(\phi_+ -\phi_{-} )$ of the initial model. In fact, the edge terms of the initial Hamiltonian $V_e(\phi_+ -\phi_{-} )$ are also added to vertex terms of the new model.

\begin{figure}[t]
\centering
\includegraphics[width=13cm,height=6cm,angle=0]{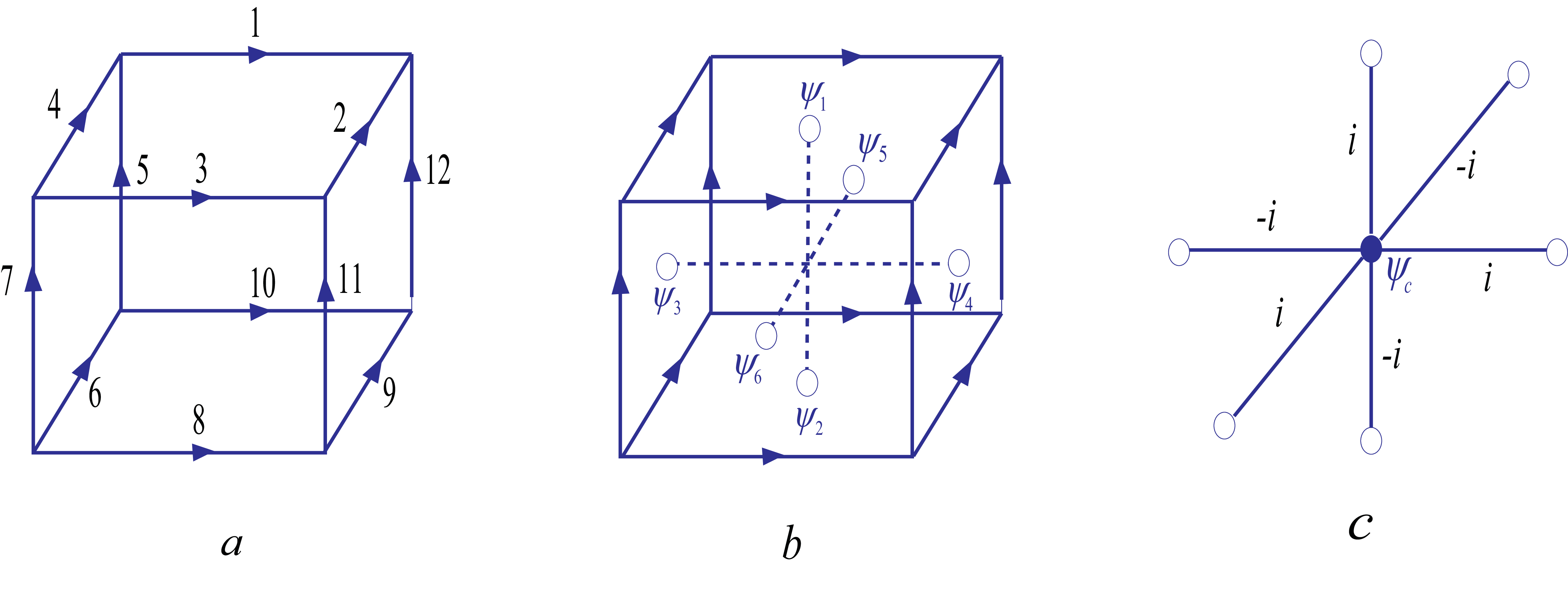}
\caption{(Color online)a) a cube of a three-dimensional lattice which is related to a 3D $U(1)$ lattice gauge theory. The field variables live in the edges of the cube. b) The plaquette variables are defined on any plaquettes of the cube which are denoted by blue circles, as an example $\psi_2 = \phi_6 -\phi_8 -\phi_9 + \phi_{10}$. c) The constraint between plaquette variables as $\psi_1 -\psi_2 -\psi_3 +\psi_4 -\psi_5 +\psi_6 =0$ is applied to the partition function by adding a new variable $\psi_c$ which is denoted by white circle.} \label{fig1-1}
\end{figure}
We can also repeat the above process for a lattice gauge theory. To this end, consider an arbitrary oriented lattice where variables $\phi_i$ live on the edges of that and the Hamiltonian is in the form of (\ref{q1}). The partition function of such a model is written in the following form:
\begin{equation}\label{a1}
\mathcal{Z}=\int D \phi~ e^{\sum_{p} J_{p} Cos(\phi_1 -\phi_2 -\phi_3 + \phi_4)}.
\end{equation}
We define plaquette variables $\psi_p =\phi_1 -\phi_2 -\phi_3 +
\phi_4 $ corresponding to each plaquette of the lattice, see
Figure (\ref{fig1-0}, left). It is clear that the new variables
$\psi_p$ are not independent and we should consider several
constraints which generally are written as $\sum_{p\in c}\pm
\psi_p=0$ where sign of field variables in this relation and form
of the constraint depends on structure of the lattice. Such
constraints can be applied to the partition function by several
delta functions and we will have:
\begin{equation}\label{a1}
\mathcal{Z}=\int D\psi ~ e^{\sum_{p}J_pCos(\psi_{p})}\prod_{c}\delta(\sum_{p\in c}\pm \psi_p).
\end{equation}
As an example in Figure (\ref{fig1-1}-a), we show a cube of a three-dimensional lattice corresponding to a 3D $U(1)$ lattice gauge theory where field variables $\phi_1 ,..., \phi_{12}$ live in the edges of that cube. As it is shown in Figure (\ref{fig1-1}-b), we define six plaquette variables $\psi_1 ,..., \psi_6$ corresponding to six plaquettes of the cube. By considering the definition of each plaquette variable $\psi_p$ in term of variables $\phi_i$, it is simple to check that a relation as $\psi_1 -\psi_2 -\psi_3 +\psi_4 -\psi_5 +\psi_6 =0$ holds. Finally, we should apply this constraint to the partition function by a delta function as $\delta (\psi_1 -\psi_2 -\psi_3 +\psi_4 -\psi_5 +\psi_6)$. 

each delta function in the above form can be written by an integral form by adding a new variable $\psi_c$ where $\delta (\psi_1 -\psi_2 -\psi_3 +\psi_4 -\psi_5 +\psi_6)=\int d\psi_c~ e^{i\psi_c(\psi_1 -\psi_2 -\psi_3 +\psi_4 -\psi_5 +\psi_6)} $, see Figure (\ref{fig1-1}-c) for a graphical notation. Generally for a lattice gauge theory with an arbitrary dimension, we can apply the above process and we have the following relation for the partition function by replacing all delta functions as exponential forms, see Figure (\ref{fig1-0}, right) for a graphical notation of the delta functions:
\begin{equation}\label{a1}
\mathcal{Z}=\int D\psi ~ e^{\sum_{p}J_pCos(\psi_{p})}\prod_{c} \frac{1}{2\pi}\int d\psi_c~ e^{i\psi_c(\sum_{p\in c}\pm \psi_p)},
\end{equation}
where we define new variables $\psi_c$ corresponding to each constraint which is denoted by $"c"$. If we apply transformations corresponding to all constraints in the partition function, it will be converted to a new model on a bipartite graph and the Hamiltonian will be in the following form:
\begin{equation}
H=\sum_{c}\sum_{p \in c}\pm i\psi_c \psi_p -\sum_{p}J_p Cos(\psi_p).
\end{equation}
Therefore similar to the classical field model, a $U(1)$ lattice gauge theory is also transformed to a classical field model with a kinetic term as $\frac{i}{2}(\psi_c -\psi_p)^2$ and the plaquette terms $Cos(\phi_1 -\phi_2 -\phi_3 + \phi_4)$ of the initial model are converted to vertex terms in the new model.

We can summarize the result of this section as the following message: The partition function of general classical field theories and $U(1)$ lattice gauge theories on arbitrary lattices are equal to the partition function of a classical field theory with a simple kinetic term as $\frac{i}{2}(\phi_i - \phi_j )^2$ on a bipartite graph. This result is the beginning of transformations that we will follow in next sections. Specially we emphasize that the Hamiltonian of the new model is still very general so that it has been defined on a complex bipartite graph and vertex terms of the new Hamiltonian are polynomial functions as $V(\phi)$, $W(\phi)$ and $Cos(\phi)$. In the next section we go toward more specification of the model by reduction of polynomial functions to a $\phi^4$ term.
\section{Reduction to $\phi^4$ term}\label{sec3}
In the previous section, we showed that the partition function of a classical field theory with a simple kinetic term in the following form is equal to the partition function of a wide set of other classical field theories and $U(1)$ lattice gauge theories:
\begin{equation}\label{b0}
H=\sum_{\la i,j \ra}\pm i\phi_i \phi_j -\sum_{i}P_i(\phi_i)
\end{equation}
where this model is defined on a complex bipartite graph and $\la i,j \ra$ refers to adjacent vertices of that graph. For brevity, we have denoted all variables by initial form $\phi_i$ and $P_i$ refers to a polynomial function which can be in different forms for different vertices.

Due to polynomial functions $P_i$, the above Hamiltonian is still very general and in this section we define an important transformation in the relation of the partition function to reduce the polynomial functions to a $\phi^4$ term. To this end, we give a simple lemma in the following form:

\textbf{Lemma}: If the polynomial function $P$ is a summation of two functions as $P(\phi)=f(\phi)+g(\phi)$, the relation of the partition function can be converted to a new one by the following form:
\begin{equation}\label{b1}
\int e^{...+P(\phi)}d\phi = \frac{1}{2\pi}\int d\psi d\psi_0 d\phi e^{...+f(\phi)+g(\psi)+i\psi_0 (\phi - \psi)},
\end{equation}
\begin{figure}[t]
\centering
\includegraphics[width=7cm,height=2cm,angle=0]{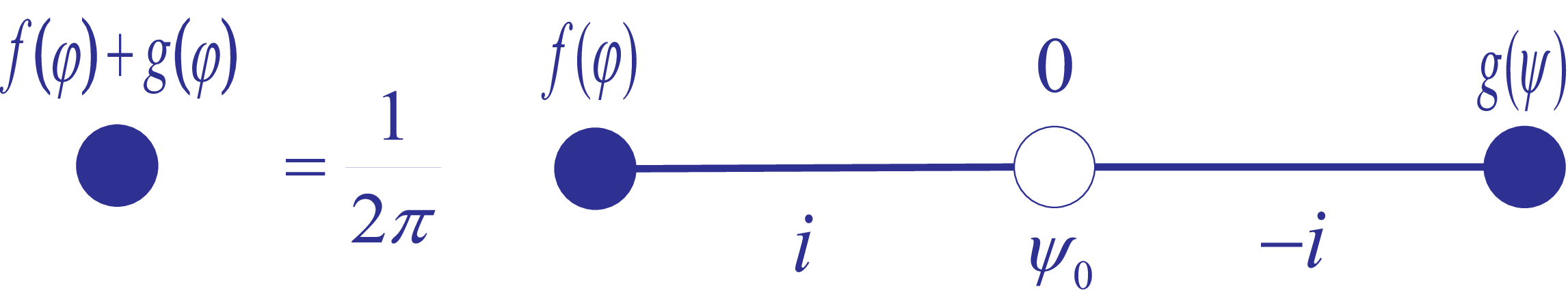}
\caption{(Color online) A vertex corresponding to a field variable can be splitted to two new vertices where the corresponding polynomial function $P$ is also splitted to functions $f$ and $g$. This rule changes the partition function by a factor $\frac{1}{2\pi}$.} \label{fig2}
\end{figure}
where $...$ in this relation refers to other terms in the Hamiltonian. To prove the above lemma, it is enough to use the integral form of the delta function as $\delta (\phi -\psi)=\int d\psi_0 e^{\psi_0 (\phi -\psi)}$. In Figure (\ref{fig2}), we also give a graphical notation for the above lemma. It shows that polynomial function $P$ on a vertex variable $\phi$ of the initial graph can be reduced to two functions $f$ and $g$ on two vertex variables of a new graph.

By lemma (\ref{b1}), we can show that each polynomial function can be reduced to $\phi^4$ term. To this end, we should take three steps as follows.

\textbf{Step 1:} Consider a polynomial function as $P(\phi)=\sum_{n=0}^{N}a_n \phi^n$ where if $N$ is infinity, it can be approximated by a big number. According to the lemma (\ref{b1}), we can reduce this function to $N$ functions on $N$ variables. Therefore, we will have a new model on a different graph where on each vertex of the graph there is only a function as $a_n \phi^n$.

\textbf{Step 2:} Although the coefficient $a_n$ generally can be any real number, we can reduce it to a real number which is very smaller than 1. To this end, we re-write the function $a_n \phi^n$ in the form of summation of $N$ terms as $\frac{a_n}{N}\phi^n+ \frac{a_n}{N}\phi^n+...+\frac{a_n}{N}\phi^n$. By this fact and the lemma (\ref{b1}), we can reduce the function $a_n \phi^n$ on a variable to $N$ functions as $\frac{a_n}{N}\phi^n$ on $N$ different variables. It is clear that if we can use a big number $N$ so that $\frac{a_n}{N}$ is smaller enough than 1.

\textbf{Step 3:} After the above steps we have a new model with functions as $t \phi^n$ where $t$ is a very small real number. To reduce such functions to $\phi^4$ terms, we start from a relation for operators and then we will give a transformation to convert such a relation to an integral relation on the classical fields. To this end, we use a well-known relation for operators where for two operators $A$ and $B$ the following Identity holds:
\begin{equation}\label{b2}
e^{tA}e^{tB}e^{-tA}e^{-tB}=e^{t^2 [A,B]+ O(t^2)},
\end{equation}
where $t$ is a real number and $[A,B]$ is the commutation of operators $A$ and $B$. If $t$ is smaller enough than 1 we can ignore from terms with order of upper than $t^2$ which are denoted by $O(t^2)$.

Suppose that $A=iP$ and $B=\frac{Q^4}{4}$ where $P$ and $Q$ are momentum and position operators respectively and $i=\sqrt{-1}$. By the fact that $[iP,\frac{Q^4}{4}]=Q^3$ and the Identity (\ref{b2}) we will have:
\begin{equation}\label{b3}
e^{itP}e^{t\frac{Q^4}{4}}e^{-itP}e^{-t\frac{Q^4}{4}}=e^{t^2 Q^3}.
\end{equation}
Although the above identity is a relation between operators, we
can convert it to an integral relation by using eigenstates of
operators $P$ and $Q$. To this end, we consider $|p\ra $ and
$|q\ra$ as eigenstates of operators $P$ and $Q$ respectively
where, from quantum mechanics, we know that $\la
p|q\ra=\frac{1}{\sqrt{2\pi}}e^{-ipq}$, we suppose $\hbar=1$. Then
we add the Identity operator $I=\int dp |p\ra \la p| = \int dq
|q\ra \la q|$ between the operators in the relation (\ref{b3}).
After these replacements we will have the following relation:
\begin{equation}\label{b4}
\frac{1}{(2\pi )^2}\int dq dq_{1} dq_{2} dp_{1} dp_{2} e^{itp_{1}}e^{t\frac{q_{1}^4}{4}}e^{-itp_{2}}e^{-t\frac{q_{2}^4}{4}+i(qp_{1} -p_{1} q_{1} +q_{1 }p_{2} -p_{2} q_{2})}|q\ra \la q_{2}|=\int dq e^{t^2 q^3}|q\ra \la q|.
\end{equation}
We can also rewrite the above relation in the following form:
\begin{equation}\label{b5}
\int dq dq_{2} \{dq_{1}dp_{1} dp_{2} \frac{1}{(2\pi) ^2} e^{itp_{1}}e^{t\frac{q_{1}^4}{4}}e^{-itp_{2}}e^{-t\frac{q_{2}^4}{4}+i(qp_{1} -p_{1} q_{1 }+q_{1} p_{2} -p_{2} q_{2})}-e^{t^{2} q^{3}}\delta(q-q_{2})\}|q\ra \la q_{2}|=0,
\end{equation}
to hold the above identity it is necessary that the following identity holds:
\begin{equation}\label{b6}
\frac{1}{(2\pi) ^2}\int dq_1dp_1 dp_2 e^{itp_1}e^{t\frac{q_{1}^4}{4}}e^{-itp_2}e^{-t\frac{q_{2}^4}{4}+i(qp_1 -p_1 q_1 +q_1 p_2 -p_2 q_2)}=e^{t^2 q^3}\delta(q-q_{2}).
\end{equation}
Since variables $p$ and $q$ are continuous variables, we can replace them with field variables $\phi$ and finally we have an interesting relation in the following form:
\begin{equation}\label{b7}
\frac{1}{(2\pi) ^2}\int d\phi d\phi_1 d\phi_2 d\phi_3 d\phi_4 e^{it\phi_1}e^{t\frac{\phi_{2}^4}{4}}e^{-it\phi_3}e^{-t\frac{\phi_{4}^4}{4}+i(\phi \phi_1 -\phi_1 \phi_2 +\phi_2 \phi_3 -\phi_3 \phi_4)}=\int d\phi e^{t^2 \phi^3}.
\end{equation}
\begin{figure}[t]
\centering
\includegraphics[width=10cm,height=1.5cm,angle=0]{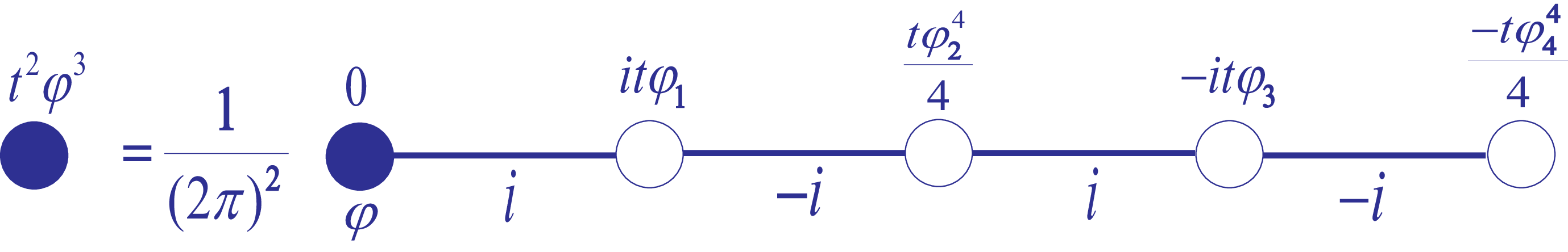}
\caption{(Color online) The partition function of a model where there is a term of $t^2 \phi^3$ on a vertex, like the left-hand figure, is equal to the partition function of a new model in the right-hand figure, where four new vertices are added and the Hamiltonian of the new model has terms of $\phi$ and $\phi^4$ on the new vertices. } \label{fig3}
\end{figure}
Now, suppose that in the partition function of our model there is a polynomial function as $t^2 \phi^3$ on a vertex of the graph. The above relation shows that function $t^2 \phi^3$ can be converted to functions $it \phi$ and $t\phi^4$ on four new vertices of a new graph. In a graphical notation it is equivalent to adding four vertices to the previous graph, see Figure (\ref{fig3}).

\begin{figure}[t]
\centering
\includegraphics[width=10cm,height=4.5cm,angle=0]{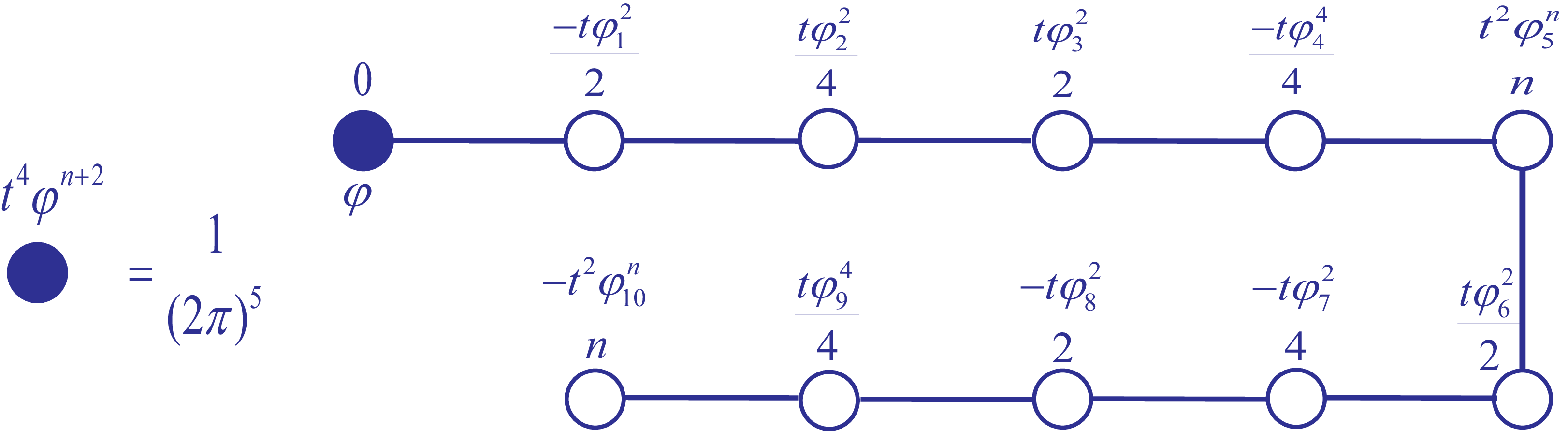}
\caption{(Color online) The partition function of a model where there is a term of $t^4 \phi^{n+2}$ on a vertex, like the left-hand figure, is equal to the partition function of a new model in the right-hand figure, where ten new variables are added and the Hamiltonian of the new model has terms of $\phi^2$, $\phi^4$ and $\phi^n$ on the new vertices. By repeating this rule we can reduce polynomial functions to $\phi^4$ term. } \label{fig4}
\end{figure}

So far we could convert function $\phi^3$ to $\phi^4$ term. By similar process we can show functions as $t \phi^n$, where
$t$ is very small, can also be converted to the $\phi^4$ term. To this end, we use another commutation relation in the following form:
\begin{equation}
[[\frac{-P^2}{2} ,\frac{ Q^4}{4}],\frac{Q^n}{n}]=Q^{n+2}.
\end{equation}
If we use this identity and apply it to relation (\ref{b2}) we will have:
\begin{equation}
e^{t^4 Q^{n+2}}=e^{t^2 [\frac{-P^2}{2} ,\frac{ Q^4}{4}]}e^{t^2 \frac{Q^n}{n}}e^{-t^2 [\frac{-P^2}{2} ,\frac{ Q^4}{4}]}e^{-t^2 \frac{Q^n}{n}},
\end{equation}
and by applying again to relation (\ref{b2}) the following identity will be derived:
$$
e^{t^4 Q^{n+2}}=e^{-t \frac{P^2}{2}}e^{t\frac{ Q^4}{4}}e^{t \frac{P^2}{2}}e^{-t\frac{ Q^4}{4}}e^{t^2 \frac{Q^n}{n}}
$$
\begin{equation}
e^{t \frac{P^2}{2}}e^{-t\frac{ Q^4}{4}}e^{-t \frac{P^2}{2}}e^{t\frac{ Q^4}{4}}e^{-t^2 \frac{Q^n}{n}}.
\end{equation}
It is simple to find integral form of this relation by adding eigenstates of operators $P$ and $Q$ similar to the previous example. We show corresponding graphical notation in Figure (\ref{fig4}). It shows that a function as $t^4 \phi^{n+2}$ on a vertex of the graph can be reduced to functions $\phi^n$ and $\phi^4$ and $\phi^2$ on ten new vertices of a new graph. If $n $ is a even number, it is clear that we can repeat this process to reduce power of $\phi$ and finally we will have only terms $\phi^2$, $\phi^4$.

Therefore if the polynomial function $P(\phi)$ in the initial Hamiltonian (\ref{b0}) is an even function like $Cos(\phi)$ for $U(1)$ lattice gauge theory, After the above transformation we have a new model on a complex graph in the following form:
\begin{equation}
H=\sum_{\la i, j \ra}\pm i\phi_i \phi_j +\sum_{i}(m_i \phi_i ^2 + J_i \phi_i ^4),
\end{equation}
where $m_i$ and $J_i$ are real numbers.

On the other hands, if $n$ is an odd number the power of $\phi$ is reduced to $\phi^2$, $\phi^3$ and $\phi^4$ but we can use relation (\ref{b7}) to convert $\phi^3$ to $i\phi$ and $\phi^4$. Therefore, if the polynomial function $P(\phi)$ in relation (\ref{b0}) is an odd function, after transformation we have a new model on a complex graph in the following form:
\begin{equation}
H=\sum_{\la i, j \ra}\pm i\phi_i \phi_j +\sum_{i}(ih_i \phi_i + m_i \phi_i ^2 + J_i \phi_i ^4),
\end{equation}
where $h_i$, $m_i$ and $J_i$ are real numbers.
\section{Transformation for decreasing dimension}\label{sec4}
In the previous sections we showed that one can map the discrete version of classical field theories and $U(1)$ lattice gauge theories on arbitrary lattices to a $\phi^4$ field model by adding many new vertices on the initial lattice in the sense that the partition function is invariant under the corresponding transformations. It is clear that the final graph after adding new vertices is a complex graph. Specially, such a graph is not a planar graph because the initial model may be in any dimension. 

In this section we give another transformation to decrease dimension of the model to two. To this end, we consider a complex graph which is derived after transformations in the previous sections. If we insert this graph on a plane, we will see many crossings between edges of the graph which shows that graph is not planar. A crossing of the graph has been shown in Figure (\ref{fig5}) where there are four variables $\phi_1$, $\phi_2$, $\phi_3$ and $\phi_4$ which interact with each other as $-i\phi_1 \phi_3 -i\phi_2 \phi_4$. Therefore, in the partition function we have an integral in the following form:
\begin{figure}[t]
\centering
\includegraphics[width=10cm,height=6cm,angle=0]{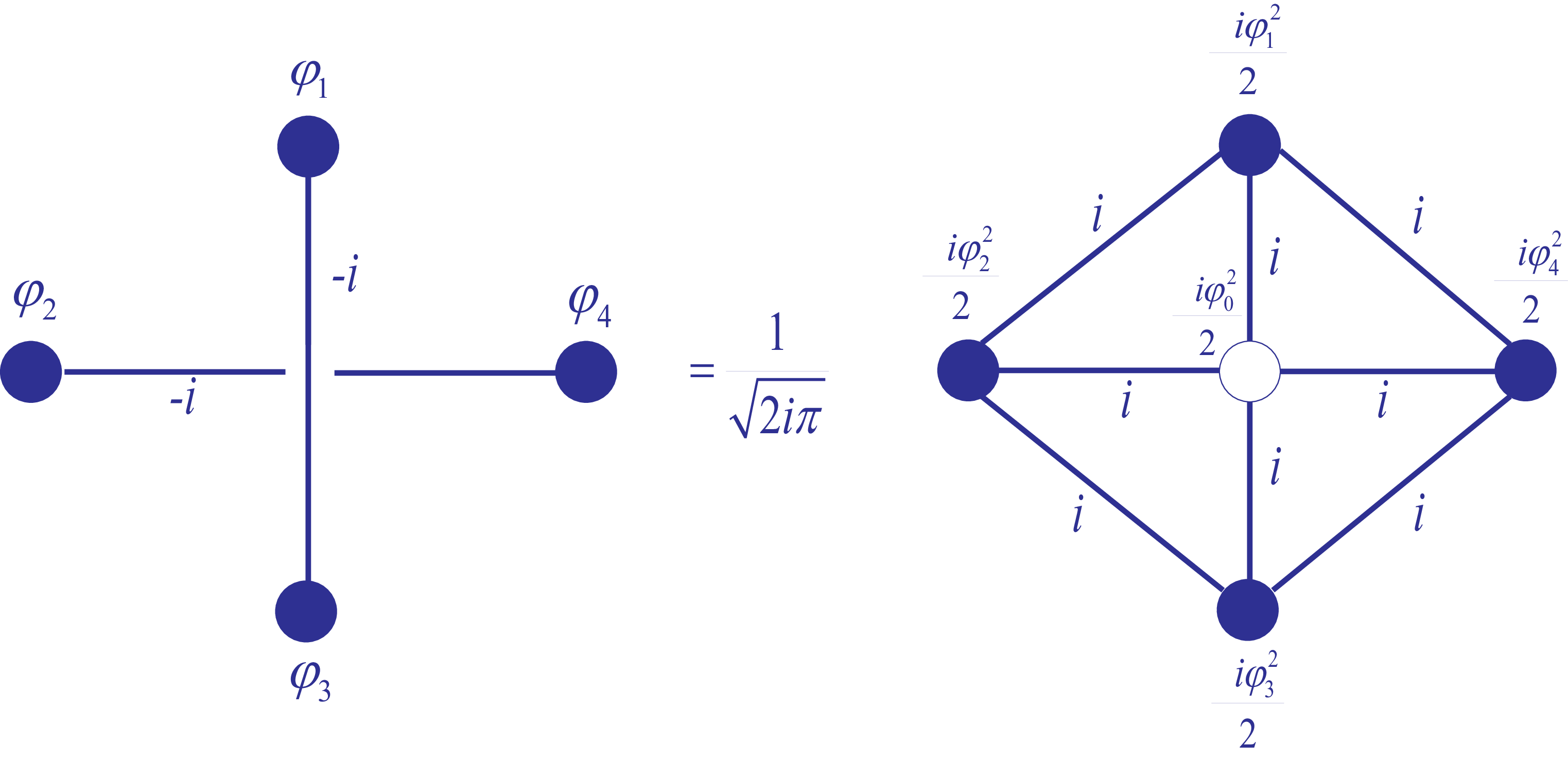}
\caption{(Color online) The partition function of a model in presence of crossing between two edges, like the left-hand figure, is equal to partition function of a new model in the right-hand figure, where a new vertex is added to remove crossing. In this way the initial graph is converted to a planar graph. } \label{fig5}
\end{figure}
\begin{equation}\label{c1}
I=\int d\phi_1 d\phi_2 d\phi_3 d\phi_4 e^{-i\phi_1 \phi_3 - i\phi_2 \phi_4}.
\end{equation}
We define a new variable $\phi_0$ and insert it into the place of crossing of the graph. Then we use a simple identity in the following form:
\begin{equation}
\int d\phi_0 e^{\frac{i}{2}(\phi_0 + \phi_1 + \phi_2 + \phi_3 + \phi_4)^2}=\sqrt{2i\pi}.
\end{equation}
We multiply this factor to the phrase (\ref{c1}) and we will have:
\begin{equation}
I=\frac{1}{\sqrt{2i\pi}}\int d\phi_0 d\phi_1 d\phi_2 d\phi_3 d\phi_4 e^{\frac{i}{2}(\phi_0 + \phi_1 + \phi_2 + \phi_3 + \phi_4)^2- i\phi_1 \phi_3 - i\phi_2 \phi_4}.
\end{equation}
After simplification we will reach to a new form of the phrase (\ref{c1}) in the following form:
\begin{equation}\label{c2}
I=\frac{1}{\sqrt{2i\pi}}\int d\phi_0 d\phi_1 d\phi_2 d\phi_3 d\phi_4 e^{i(\phi_1 \phi_2 + \phi_2 \phi_3 +\phi_3\phi_4 +\phi_4\phi_1) + i\phi_0 \sum_{i=1}^4 \phi_i +\frac{i}{2}\sum_{i=0}^4 \phi_{i}^2},
\end{equation}
The transformation from phrase (\ref{c1}) to phrase (\ref{c2}) has been shown graphically in Figure (\ref{fig5}). It shows that a crossing in the initial graph can be removed by adding a new variable and some changes in the pattern of connectivity of the initial graph.

Finally, after removing all crossings in the initial graph we have a new model on a planar graph in the following form:
\begin{equation}\label{d1}
H=\sum_{\la i,j \ra}\pm i\phi_i \phi_j +\sum_{i}(i h_i \phi_i + m_i \phi_i ^2 + J_i \phi_i ^4),
\end{equation}
where $h_{i}$ is a real number which was necessary for odd functions. Furthermore, Since complex function $\frac{i}{2}\phi_i^2$ is necessary for reduction of dimension, $m_{i}$ can be a complex number while $J_i$ is still a real number.
\section{Matching to the 2D square lattice}\label{sec5}
After transformation of the initial model to a $\phi^4$ theory on a planar graph in the previous section, we are ready to complete the result by matching the model on a 2D square lattice. The Hamiltonian (\ref{d1}) is defined on a complex planar graph. In order to match such a graph to a square lattice, there are three points which should be considered. We explain these points in three steps:

\textbf{Step 1: Reducing the degree of vertices}
\begin{figure}[t]
\centering
\includegraphics[width=8cm,height=2.5cm,angle=0]{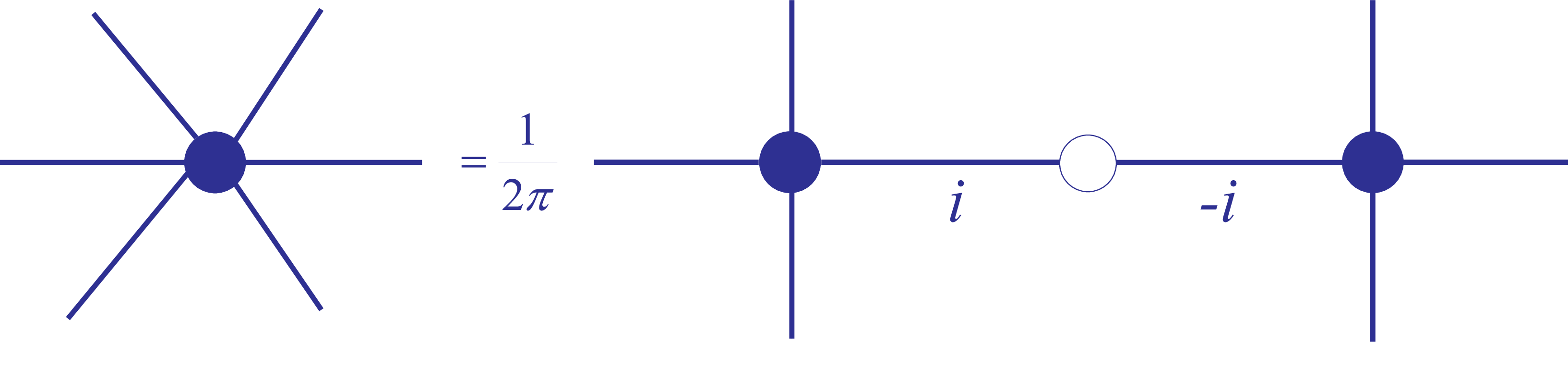}
\caption{(Color online) The partition function of a model where there is six links for a vertex, like the left-hand figure, is equal to the partition function of a new model in the right-hand figure, where the initial vertex is splitted to two new vertices with lesser links. If the degree of the original vertex is upper than six, it is again possible to reduce the degree of all vertices to four by repeating this rule. } \label{fig6}
\end{figure}
The most important problem for matching the graph on a square lattice is to reduce the degree of vertices. In a square lattice, the degree of each vertex is four while in the our model (\ref{d1}) the degree of each vertex may be any arbitrary numbers. We use the lemma (\ref{b1}) to reduce the degree of a vertex. For example, according to lemma (\ref{b1}) as it is shown in Figure (\ref{fig6}), we can split a vertex which has six links with other vertices of the lattice to two new vertices where there are only three connections for each one of them. For the vertices of the graph with higher degrees it is enough to repeat this rule for more times to finally reduce the degree to four.

\textbf{Step 2: Link insertion}
Second problem is that we should match all edges of the final graph to edges of the square lattice. To this end, it is necessary to insert several vertices on the edges of the graph to match it to the vertices of the square lattice. We use a simple rule for adding vertices on any links of the graph in the following form:
\begin{equation}
\int d\phi_1 d\phi_2 e^{i\phi_1 \phi_2 - \frac{i}{2}\phi_1^2 - \frac{i}{2}\phi_2^2}= \frac{1}{\sqrt{2\pi i}}\int d\phi_0 d\phi_1 d\phi_2 e^{i\phi_0 \phi_1 -i\phi_0 \phi_2 + \frac{i}{2}\phi_0^2}.
\end{equation}
\begin{figure}[t]
\centering
\includegraphics[width=8cm,height=1.5cm,angle=0]{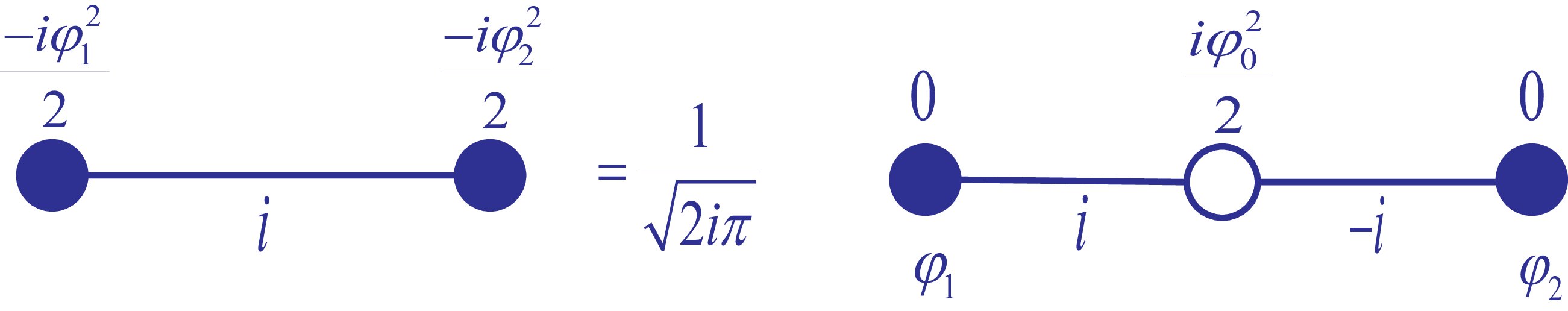}
\caption{(Color online) Adding a new vertex to a link in the left-hand figure changes the partition function only by a complex factor $\frac{1}{\sqrt{2i\pi}}$ and also changes polynomial functions on variables as shown the right-hand figure.} \label{fig7}
\end{figure}
This relation has been shown in Figure (\ref{fig7}) in a graphical notation. To prove this relation it is enough to complete the square as $(\phi_0 -(\phi_2 -\phi_1))^2$ in the right-hand of the relation and perform integration on $\phi_0$. In this way, we can add enough number of vertices to completely match the graph to the square lattice.

\begin{figure}[t]
\centering
\includegraphics[width=10cm,height=4cm,angle=0]{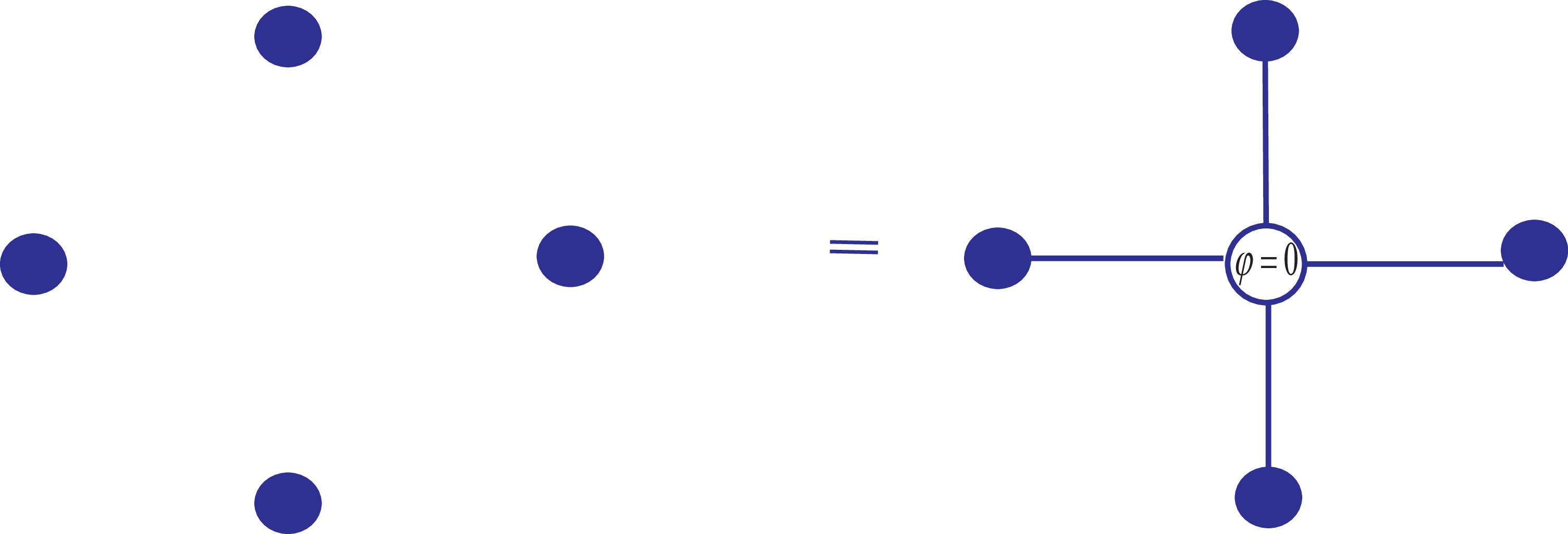}
\caption{(Color online) Adding a new vertex to each empty point of the lattice, where the corresponding field $\phi$ is fixed to zero, does not change the partition function.} \label{fig8}
\end{figure}

\textbf{step 3: Face insertion}
Last step for complete matching to a square lattice is that there is also some empty points on square lattice which should be filled by new variables. To this end, we can add a new variable $\phi$ in each point of the square lattice and fix its value to zero. As it is shown in Figure(\ref{fig8}), such a transformation does not lead to any extra terms in the partition function.
In this way by proper combination of the above steps, one will be able to convert the original graph to a 2D rectangular lattice.

\section{completeness of 2D classical $\phi^4$ theory}\label{sec6}
In the previous sections we gave few steps to convert a general field theory to a new model on a square lattice in the following form:
\begin{equation}
H=\sum_{\la i , j \ra} \pm i \phi_i \phi_j + \sum_{i}(i h_i \phi_i +m_i \phi_i ^2 +J_i \phi_i ^4),
\end{equation}
by a simple transformation it is possible to change this model to standard $\phi^4$ field theory. To this end, we use this fact that $i \phi_i \phi_j =-\frac{i}{2}(\phi_i - \phi_j)^2 + \frac{i}{2} \phi_i ^2 + \frac{i}{2}\phi_j ^2$ and after replacement of this relation to the above Hamiltonian we will have:
\begin{equation}\label{t1}
H=\sum_{\la i , j \ra} \pm \frac{1}{2} (\phi_i - \phi_j)^2 + \sum_{i}(i h_i \phi_i +m_i \phi_i ^2 +J_i \phi_i ^4),
\end{equation}
where we absorb $ \pm \frac{i}{2} \phi_i ^2$ in the term $m_i \phi_i ^2$. In this way, our complete model is a classical $\phi^4$ field model on a square lattice where coupling constants of the model can be complex numbers while $J_i$ are real numbers.

There is also another important point that we should consider about our results. In fact, for transformation of the initial model to a complete model as (\ref{t1}) we added several new vertices to the original lattice. such a work would not be efficient if the number of added vertices were as an exponential function of number of vertices in the original model. But it is simple to show that in all transformations that we gave, the number of added vertices were limited so that total number of added vertices certainly will be a polynomial function of initial number of vertices. Furthermore, we emphasize that although some our steps in order to convert various models to a $\phi^4$ field theory are performed with approximations, it is possible to perform transformations with arbitrary precision.
\section{Discussion}
One of the most important problems in the theory of completeness
in SM is to find an explicit pattern of couplings in a specific complete model
which generates the partition function of different SM models. In
this paper, we took an important step toward such a problem for
the completeness of $2D$ classical $\phi^4$ field theory. We gave
a step by step proof which shows how the partition function of a
discrete version of classical field theories and also $U(1)$
lattice gauge theories on arbitrary lattices with arbitrary
dimensions can be derived from a 2D classical $\phi^4$ field
theory. Furthermore, we believe that our approach has this
advantage that is more general than the previous proof. Therefore,
it leads to a general approach which can be used by various
communities of physicists for future researches on similar
problems.
\section{acknowledgement}
M. H. Zarei would like to thank M. M. Golshan and A. Poostforush because of very good comments and their helps for editing paper.

\end{document}